\documentstyle[12pt]{article}
\newcommand{\nc}{\newcommand}
\nc{\varPhi}{\Phi} \nc{\drm}{{\rm d}} \nc{\bi}{\bibitem} \nc{\til}{\tilde}
\nc{\Lc}{{\cal L}} \nc{\om}{\omega} \nc{\bb}{\begin{equation}}
\nc{\ee}{\end{equation}} \nc{\vecna}{\mbox{\boldmath $\nabla$}}
\nc{\ga}{\gamma} \nc{\erm}{{\rm e}} \nc{\ib}{{\bf i}}
\nc{\sig}{\sigma} \nc{\longra}{\longrightarrow} \nc{\al}{\alpha}
\nc{\vare}{\varepsilon} \nc{\C}{I\!\!\!C} \nc{\hb}{\hbar}
\nc{\1}{1\!\!1} \nc{\lan}{\langle} \nc{\ran}{\rangle}
\nc{\Om}{\Omega} \nc{\R}{I\!R} \nc{\h}{\hspace*{0.5 cm}} \nc{\wide}{\widehat}
\nc{\Psit}{\widetilde{\Psi}} \nc{\Rt}{\widetilde{R}}
\nc{\psib}{\overline{\psi}} \nc{\psit}{\widetilde{\psi}} \nc{\bt}{\beta}
\nc{\und}{\underline} \nc{\sta}{\stackrel} \nc{\cent}{\centerline}
\nc{\vs}{\vspace*} \nc{\zb}{\overline{\psi}} \nc{\doz}{\dot{\psi}}
\nc{\ddoz}{\ddot{\psi}} \nc{\dozb}{\dot{\overline{\psi}}}
\nc{\ddozb}{\ddot{\overline{\psi}}} \nc{\dox}{\dot{x}} \nc{\dopi}{\dot{\pi}}
\nc{\dopsi}{\dot{\psi}} \nc{\Vbf}{\mbox{\boldmath $V$}}
\nc{\sbf}{\mbox{\boldmath $s$}} \nc{\Sigbf}{\mbox{\boldmath $\Sigma$}}
\nc{\ombf}{\mbox{\boldmath $\omega$}} \nc{\gabf}{\mbox{\boldmath $\gamma$}}
\nc{\omp}{\omega_{{\rm \psi}}}
\nc{\vbf}{\mbox{\boldmath $v$}} \nc{\impulse}{\mbox{\boldmath $p$}}
\nc{\vo}{{\wide {\vbf}}} \nc{\po}{{\wide {p}}} \nc{\Po}{{\wide {P}}}
\nc{\impo}{\wide {\impulse}} \nc{\ii}{\`{\i \ }} \nc{\ug}{\; = \;}

\begin{document}

\baselineskip 0.7cm

\cent{\bf THE SPINNING ELECTRON: HYDRODYNAMICAL REFORMULATION,}

\cent{\bf AND QUANTUM LIMIT, OF THE BARUT--ZANGHI THEORY$^{(\dagger)}$}

\footnotetext{$^{(\dagger)}$ Work partially supported by INFN, MURST, CNR,
and by CNPq.}

\vs{0.5cm}

\begin{center}

{Giovanni SALESI

{\em Dipart.$\!$  di Fisica, Universit\`a Statale di Catania, 57 Corsitalia,
95129--Catania, Italy;

and INFN, Sezione di Catania, Catania, Italy}

\vs{0.2cm}

and

\vs{0.2cm}

Erasmo RECAMI

\vs{0.1cm}

{\em Facolt\`a di Ingegneria, Universit\`a Statale di Bergamo,
24044--Dalmine (BG), Italy;

INFN, Sezione di Milano, Milan, Italy; \ and

Dept. of Applied Math., State University at Campinas, Campinas, S.P.,
Brazil.}}

\end{center}

\vs{1.0cm}

{\bf Abstract \ --} \  One of the most satisfactory pictures
for spinning particles is the Barut-Zanghi (BZ) classical theory for
the  relativistic extended-like  electron, that relates spin to
{\em zitterbewegung} (zbw). The BZ motion equations constituted the starting
point for recent works about spin and electron structure, co-authored by us,
which adopted the Clifford algebra language. \
This language results to be actually suited and fruitful for a
hydrodynamical re-formulation
of the BZ theory.  Working out, in such a way, a ``probabilistic fluid'', we are
allowed
to re-interpret the original classical spinors as quantum wave-functions for
the electron.  \ Thus, we can pass to ``quantize" the BZ theory  employing
this time the tensorial language, more popular in first-quantization. \
``Quantizing'' the BZ theory, however, does {\em not}
lead to the Dirac equation, but rather to a non-linear, Dirac--like
equation, which can be regarded as the actual ``quantum limit'' of
the BZ classical theory. \
Moreover, an original variational approach to the the BZ
probabilistic fluid shows that it is a typical ``Weyssenhoff fluid'', while
the Hamilton-Jacobi equation (linking together mass, spin and zbw frequency)
appears to be nothing but a special case of de Broglie's famous
energy-frequency relation. \ 
Finally, after having discussed the remarkable correlation between the 
gauge transformation U(1) and a general rotation on the spin plan, \
we clarify and comment on the two-valuedness nature of the fermionic
wave-function, and on the parity and charge conjugation transformations.

\newpage

\section{The Barut--Zanghi theory for the classical spinning electron}

\h Since the works by Compton,$^{[1]}$ Uhlenbeck and Goudsmith,$^{[2]}$ and
Frenkel,$^{[3]}$  many
classical models of spin and classical theories of the electron have been
investigated for about seventy years.$^{[4-6]}$ For instance, Schr\"{o}dinger's
suggestion$^{[7]}$ that the electron spin was related to
{\em zitterbewegung} (zbw) did originate a large amount of
subsequent work, including Pauli's.
The zbw is actually the spin motion, or ``internal'' motion ---since it is
observed
in the center-of-mass frame (CMF),--- expected to exist for spinning
particles. It arises because the motion of the electrical charge
does {\em not} coincide with the motion of the particle CM.
In the Dirac theory, indeed,
the velocity and impulse operators $\vo$ and $\impo$ are not parallel:
$$
\vo\neq\impo/m\; .
\eqno{(1)}
$$
So a zbw {\em is} to be added to the usual
drift, translational, or ``external'', motion of the CM, $\,\impulse/m$
(which is the only one to occur in the case of scalar particles).
In Barut--Zanghi's (BZ) theory,$^{[8,9]}$ the classical electron was
actually characterized, besides by the usual pair of conjugate variables
$(x^\mu, p^\mu)$, by a second pair of conjugate classical {\em spinorial}
variables $(\psi, \overline{\psi})$, representing internal degrees of freedom,
which are functions of the (proper) time $\tau$  measured in the
electron CMF; the CMF being the one ---let us recall--- in which
$\impulse = 0$ at any instant of time.  \ Barut and Zanghi introduced,
namely, a classical lagrangian that in the free case (i.e., when the
{\em external}
 electromagnetic potential is $A^\mu = 0$) writes $[c=1]$
$$
\Lc=\frac{1}{2} i \lambda (\dot{\overline{\psi}}\psi - \overline{\psi}\dot{\psi}) +
p_\mu(\dot{x}^\mu - {\overline{\psi}}\ga^\mu \psi) \; ,
\eqno{(2)}
$$
where $\lambda$ has the dimension of an action, and $\psi$ and
$\overline{\psi}\equiv \psi^{\dag} \ga^0$ are ordinary
${\rm {\C}}^4$--bispinors, the dot meaning derivation with respect to $\tau$.
\ The four Euler--Lagrange equations, with $-\lambda=\hbar=1$,
yield the following motion equations:

\

$\hfill{\displaystyle\left\{\begin{array}{l}
\dot{\psi}+ i p_\mu \ga^\mu \psi=0\\

\dot{x}^\mu= \overline{\psi}\ga^\mu \psi\\

\dot{p}^{\mu}=0 \; ,
\end{array}\right.}
\hfill{\displaystyle\begin{array}{r}
(3{\rm a}) \\ (3{\rm b}) \\ (3{\rm c}) \end{array}}$

\

besides the hermitian adjoint of eq.(2a), holding for $\overline{\psi}$. \
From eq.(2) one can also see that
$$
H \equiv p_{\mu} v^{\mu} = p_{\mu} \overline{\psi} \ga^{\mu} \psi
\eqno {(4)}
$$
is a constant of the motion [and precisely is the energy in the
CMF].$^{[8]}$ \ Since $H$ is the BZ hamiltonian in the CMF, we can
suitably set \ $H=m$, \ where $m$ is the particle rest-mass.
The general solution of the equations of motion (3) can be shown to be:
$$
\psi(\tau)=\left[\cos (m\tau)- i \frac{p_\mu \ga^\mu}{m}
\sin (m\tau)\right]\psi(0) \  ,
\eqno{\rm(5 a)} $$             

$$
{\overline{\psi}}(\tau)={\overline{\psi}}(0)\left[\cos (m\tau) + i \frac{p_\mu \ga^\mu}{m}
\sin (m\tau)\right] \  ,       
\eqno{\rm(5 b)} $$

with $p^\mu=$ constant; \ $p^2=m^2$; \ and finally:
$$
\dot{x}^\mu\equiv v^\mu=\frac{p^\mu}{m}+\left[\dot{x}^\mu(0)-\frac{p^\mu}{m}
\right]
\cos(2m\tau)+\frac{\ddot{x}^\mu}{2m}(0)\sin (2m \tau) \ .
\eqno {\rm(5 c)} $$            

This general solution exhibits the classical analogue of the zbw: in fact, the
velocity $v^\mu$ contains the (expected) term $p^\mu/m$ plus a term describing
an oscillating
motion with the characteristic zbw frequency $\om=2m$. \ The velocity
of the CM will be given by $p^\mu/m$. Let us explicitly observe
that the general solution (5c)
represents a helical motion in the ordinary 3-space: a result that has been met
also by means of other, alternative approaches.$^{[8]}$

\h Notice that, instead of adopting the variables $\psi$ and $\overline{\psi}$, we can
work in terms of the ``spin variables'', i.e., in terms of the set of dynamical
variables
$$
x^\mu \, , \ p^\mu \, ; \ v^\mu \, , S^{\mu \nu}
\eqno{(6)}
$$
where
$$
S^{\mu \nu} \equiv {i \over 4} \, \overline{\psi} [\ga^\mu, \ga^\nu] \psi \;
\eqno {(7)}
$$
is the particle spin tensor;
then, we would get the following motion equations:
$$
\dot{p}^\mu=0 \ ; \ \ v^\mu=\dot{x}^\mu \ ; \ \ \dot{v}^\mu=4 S^{\mu
\nu}p_{\nu} \ ; \ \ \dot{S}^{\mu \nu}= v^\nu p^\mu - v^\mu p^\nu \ .
\eqno {(8)}
$$
The last equation expresses the conservation of the total angular momentum
$J^{\mu\nu}$, sum of the orbital angular momentum $L^{\mu\nu}$ and
of $S^{\mu\nu}$:
$$
{\dot {J}}^{\mu\nu} = {\dot {L}}^{\mu\nu} + {\dot {S}}^{\mu\nu} = 0 \; ,
\eqno{(9)}
$$
being ${\dot {L}}^{\mu\nu} = v^\mu\pi^\nu - v^\nu\pi^\mu$ from the very
definition of $L$.


\h Furthermore, in the last two refs.[9] it was found that free polarized 
particles (i.e., with the spin projection $s_z$  equal to
$\pm{1\over 2}$) are endowed with internal {\em uniform circular} motions, and
vice-versa. \
In such a way, the only values for $s_z$ corresponding to {\em classical} uniform
motions in the CM frame (just the ones expected for free particles) belong
to the discrete {\em quantum} spectrum $\pm{1\over 2}$. \
It was also therein noticed that such zbw motions are the only motions for which the
square of the 4-velocity is constant in time. The radius of the orbit in the CM frame
was found to be equal to $|\Vbf|/2m$ \ (quantity $\Vbf$ being the orbital 3-velocity), which,
in the special case of a {\em light-like} zbw, turns out to be equal to half the Compton
wave-length. \
Subsequently the Euler--Lagrange equations were generalized to the
case of an electron in an electromagnetic field, and
the analytical solutions of the motion equation in the special case of
uniform magnetic field were written down.


\h The BZ theory has been recently studied also
in the lagrangian and hamiltonian symplectic formulations, both in
flat and in curved spacetimes.$^{[8]}$\\

\section{Hydrodynamics and quantum limit of the BZ theory}

\h The {\em Multivectorial} or {\em Geometrical} Algebras are essentially due to
the work of great mathematicians of the nineteenth century as
Hamilton (1805--1863), Grassmann (1809--1977) and Clifford (1845--1879).
Starting from the sixties, some
authors, and in particular Hestenes,$^{[10]}$ sistematically
studied the interesting physical applications of such algebras, and
especially of the {\em Real Dirac Algebra}, often renamed as {\em Space-Time
Algebra} ${\rm {\R}}_{1,3}$ (STA).$^{[10]}$
In microphysics, we can meet applications for the case of space-time
[O(3), Lorentz] transformations, gauge [SU(2), SU(5), strong and electroweak
isospin] transformations, chiral [SU(2)$_L$] transformations, Maxwell
equations, magnetic monopoles$^{[11]}$, and so on.
But the most rigorous application is probably the formal and conceptual
analysis of the geometrical, kinematical and {\em hydrodynamical} content of
the Pauli and Dirac equations, performed by means
of the Real Pauli and Dirac Algebras, respectively.
We are now going to see that, not only for the Dirac probabilistic ``fluid'',
but also for the BZ probabilistic fluid, the Clifford Algebras language
results to be actually
suited and fruitful.  We shall obtain in the next Sections
the hydrodynamical (often said {\em local} or {\em field}) formulation in
the STA
of the BZ theory. In this Section we first get the field equations for
the BZ electron; then, by translating from STA into standard algebra,
we work out a ``quantization'' of the BZ
theory. As a consequence, we arrive to a non-linear Dirac--like
wave-equation, which can be actually regarded as the ``quantum limit'' of
the BZ classical theory.
Finally, in the next Section, a variational STA approach
to the the BZ ``fluid'' will lead us to the conclusion that it is a typical
``Weyssenhoff fluid''.$^{[12]}$\\

\h The {\em translation} of the single terms of lagrangian (1) into the STA
language can be performed as follows:$^{[9]}$

\

\hfill{$\begin{array}{lcl} {1 \over 2}i(\dozb\psi - \psib\doz) & \longra & \lan
\psit \dopsi \ga_1 \ga_2 \ran_0  \\ \pi_{\mu}(\dox^{\mu} - \zb \gabf^{\mu} \psi)
&  \longra & \lan \pi(\dox - \psi \ga_0 \psit ) \ran_0  \\ e A_{\mu} \zb
\gabf^{\mu} \psi & \longra & e \: \lan A \psi \ga_0 \psit \ran_0 \; , \end{array}$
\hfill}

\

\noindent where, as usual, $\psi$ indicates the so-called {\em Dirac real
spinor}, whilst $\lan \;\;\; \ran_0\;$ represents the so-called
{\em scalar part of the Clifford product}.$^{[9,10]}$
\ In such a way, our lagrangian becomes:
$$
\Lc \; = \; \lan \psit \dopsi \ga_1 \ga_2 \: + \: \pi(\dox -
\psi \ga_0 \psit) \: + \: eA \psi \ga_0 \psit \ran_0 \; .
\eqno{(10)}
$$
\noindent The Eulero--Lagrange equations now read:

\

\hfill{$\dopsi \ga_1 \ga_2 + \pi \psi \ga_0 \; = \; 0 $
\hfill} (11a)

\hfill{$\dox \; = \; \psi \ga_0 \psit $
\hfill} (11b)

\hfill{$\dopi \; = \; e F \cdot \dox $
\hfill} (11c)

\

\noindent where \ $F \equiv \partial \wedge A$ \ is the
electromagnetic field {\em bivector}.

\h In view of a quantum interpretation of the BZ theory, we need a
formulation and analysis of it in terms of spinors, which be no longer
functions of $\tau$, but instead of $x^{\mu}$ (spinorial {\em
fields} $\psi (x)$). At the same time the BZ theory gets a sort of
conceptual, and not merely formal, ``extension''. In fact, it will result
to describe a ``fluid'' which admits a {\em probabilistic
interpretation}; its integral stream-lines$^{\#}$
\footnotetext{$^{\#}$ We refer to {\em congruence} of world-lines.}
coinciding with the single (semi)-classical world-lines (up to now
parametrized in terms of $\tau$) of the point-like electron charge.
Beside the spinorial field $\psi (x), \psib (x)$, we shall meet also the
fields $v(x), \; p(x), \; \sbf (x), \; S^{\mu\nu}(x)$, which will replace the
corresponding functions of $\tau$.
And the spinorial field $\psi(x)$ will be such that its {\em restriction}
${\psi(x)}_{\mid \sig}$ to the
world-line $\sig$ (along which the particle moves) coincides with $\psi(\tau)$.
At the same time, the velocity distribution $V(x)$ is required to be
such that its restriction ${V(x)}_{\mid \sig}$ to the world-line $\sig$
results to be just the ordinary 4-velocity $v(\tau)$ of the considered
particle. Therefore,
for the tangent vector along any line $\sig$ the relevant relation holds:
$$
\frac{\drm}{\drm \tau} \equiv \frac{\drm x^\mu}{\drm \tau} \frac{\partial}{\partial
x^\mu}\equiv {\dot{x}}^\mu \partial_\mu  \ . \eqno{(12)}
$$
Eq.(12) is nothing but the direct relativistic extension of the
well-known  equation found in the non-relativistic theory of fluids, and
linking each other the ``eulerian'' (or ``spatial'') and the ``lagrangian''
(``material'') velocities:
$$
\frac{\drm}{\drm t} \equiv \partial_t + \vbf\cdot\vecna\; .
$$
Inserting the total derivative (12) into the Euler--Lagrange equation (11a),
we get:

\

\hfill{$v \cdot \partial \psi \ga_1 \ga_2 + \pi \psi \ga_0 \; = \; 0 \; ,$
\hfill} (13)

\

\noindent and, it being $\dot{x} = \psi\ga_0\psit$ because of eq.(11b), we
finally obtain the following noticeable equation:

\

\hfill{$(\psi \ga_0 \psit) \cdot \partial \psi \ga_1 \ga_2 + \pi \psi \ga_0 \; =
\; 0 \; .$
\hfill} (14)

\

\h Equation (14) expresses the ``field'' content of the BZ theory.
Incidentally, let us notice that, differently from eq.(11a), equation (14)
can be valid a priori even for massless spin $1 \over 2$ particles, since
the CMF proper time does not enter it any longer.
\ We see also that, since the restriction of $\psi (x)$ to the world-line $\sig$
coincides with $\psi(\tau)$, the velocity field
$V(x) \equiv \psi(x)\ga_0\psit(x)$
results to be endowed with the same zbw as found for $v(\tau)$ (with the
oscillation $m\tau$ suitably replaced by the equivalent quantity $p\cdot x$,
in the free-case).\\

\h The ``quantum re-formulation'' of the classical BZ theory
essentially consists in:

i) re-interpreting the spinor field $\psi (x)$ in eq.(14) as the proper
{\em
wave-function} for the spinning particle (or also, in a second quantization
formulation, as the {\em creation operator} ${\wide {\psi}}$ for the particle
{\em quantum field});

ii) re-interpreting the originary BZ theory as the {\em classical limit} of the
non-linear {\em quantum wave-equation} we are going to get (see below).

\h Obviously, for the probabilistic re-interpretation i), we must consider
the bilinear quantity $\psi\psit$ as a {\em probability density}.
\ Let us now translate eq.(14) into the ordinary tensorial language (limiting
ourselves for simplicity to the free case),
so that $\psi$ does loose the direct geometrical meaning owned within the STA,
playing on the contrary the customary r\^{o}le of wave-function in quantum
mechanics.
By means of the usual corrispondence rules linking the STA and the standard
algebra,$^{[9,10]}$

\

\hfill{$\psi\ga_0\psit\ga_\mu \; \longra \; \overline{\psi}\ga_\mu\psi $
\hfill} (15a)

\hfill{$\partial_\mu\psi\ga_2\ga_1 \; \longra \; i\partial_\mu\psi $
\hfill} (15b)

\hfill{$p\ga_0\psi \; \longra \; m\psi $
\hfill} (15c)

\

\noindent we straightly get the following {\em non-linear} Dirac--like
equation, the ``quantum limit'' of the BZ theory:
$$
i\overline{\psi}\ga^\mu \psi \partial_\mu \psi \; = \; m\psi\ .
\eqno{(16)}
$$
\h Let us remark that the non-linearity with respect to $\psi$ was already
present in the original Euler--Lagrange equations, because of eq.(11b) in
which the bilinear quantity $\psi \ga_0 \psit$ first appeared. \
``Quantizing'' the BZ theory, therefore, does {\em not}
lead to the Dirac equation, but rather to such a non-linear, Dirac--like
equation.\\

\section{Variational approach to the BZ fluid}

\h By means of Clifford algebras, we can now work out, in a variational
(lagrangian) context, the set of equations
necessary for a complete description of the BZ--fluid hydrodynamics. \
In STA a Dirac spinor may be written as follows:
$$
\psi \; = \; \sqrt{\rho}\,\erm^{{1\over 2}\bt\gamma_5}
R_0\,\erm^{\hbar\ga_1\ga_2\varphi}\;
$$
\h The scalar $\rho$, the ``proper density'', works as a normalization factor;
the
scalar $\bt$ is the ``Takabayasi angle''; $\ga_{5}$ indicates the
pseudoscalar unit of the STA; and the scalar $\varphi$ may be considered as a
``generalized''
spinor phase. The bivector $R_0 \equiv R_0(S)$ depends exclusively on the
bivector $S$. The constant bivector $\hbar\ga_2\ga_1$ is generally associated to
``the oriented {\em spin plane}'', so that in a generic frame we have
$S = {1\over 2}R_{0}\hbar\ga_2\ga_1\Rt_{0}$ , where by $\Rt_{0}$ we indicate
the result of the
``reversion'' operation on ${R_0. \;}^{[10]}$

\h After some algebra we obtain:
$$
\Lc \; = \; \rho\cos\bt(\Om\cdot S - {\dot \varphi}) + p\cdot(\dox - \rho v)
\eqno{(17)}
$$
where (as functions of the chosen lagrangian variables $\rho, \varphi,
\bt, R_0$) the following relations hold
$$
\Om \equiv 2{\dot R}\Rt \equiv 2{\dot {R_0}}\Rt_0
\qquad \qquad \Om\cdot S \equiv {\dot {R_0}}\hbar\ga_2\ga_1\Rt_0
\qquad \qquad v \; \equiv \; R_0\ga_0\Rt_0\; ,
\eqno{(18)}
$$
$\Om$ being the so-called {\em angular velocity} bivector.$^{[10]}$

\h Notice that lagrangian (17) results to be the sum of two expressions which
vanish (yielding in this way the equations of the motion), both multiplied by
suitable ``Lagrange multipliers'' ($\rho \cos\bt$ and  $p$).

\h Let us now take the variations with respect to $\rho, \varphi, \bt, R_0$:
we shall obtain in the same order the following equations:

\

\h A) {\em The ``Hamilton--Jacobi'' equation:}

\

\cent{$p\cdot v \ug \Om\cdot S\cos \bt \; .$}

\

\noindent As it is easy to verify, it holds$^{[9]}$ that $\bt = 0$ for
the solution (5a) of the BZ theory (suitably translated into the STA), so that
we may take $\cos \bt = 1 .$ In such a way, our first equation shows the {\em
kinematico--geometrical content of the} celebrated {\em de Broglie's relation}
$E = \hbar\omega$:
$$
H_{\rm CMF} = m = \Om\cdot S \equiv \ombf\cdot\sbf = {1\over 2}\hbar\om =
\hbar\omp \; ,
\eqno{(19)}
$$
were we indicated by $\om = 2m$ the zbw motion frequency [that is, the
frequency
appearing in the motion equation (5c)]; and by $\omp = m$ the frequency of the
spinor $\psi$, appearing in the solution (5a) [and corresponding, after
our  ``quantum re-interpretation'', to the frequency of the wave-function].

\h A wave-plane is a mathematical device found in the quantum formalism which is
not endowed with a direct, intuitive physical meaning; and the Planck constant,
appearing through the whole quantum theory, is not deduced from a
physical context, but is required {\em a priori} and inserted ``by hand''.
{\em It is therefore noticeable the possibility of replacing, in the de Broglie
relation, the wave-plane frequency by the zbw motion frequency, as well as the
Planck constant by spin}.
\ Starting from the interpretation of $\hbar /2$ as actually meaning
$|\sbf|$, one of the present authors has recently deduced the so-called ``quantum
potential'' of the Madelung fluid as being the kinetical energy of the
zbw.$^{[13]}$ \
Let us here only notice that expressing  the mass $m$ in the form
$\ombf\cdot\sbf$ seems to denounce the origin of the particle mass
as due to a sort of ``rotational kinetic--energy''.$^{[10]}$

\

\h B) {\em The continuity equation:}

\

\cent{${\dot \rho} \ug 0 \; .$}

\

\noindent As expected, along a stream-line the flux density is constant in time.

\

\h C) {\em A correlation between spinorial phase and angular velocity:}

\

\cent{${\dot\varphi} \ug \Om\cdot S \equiv {1\over 2}\om$}

\

\noindent By integrating this equation, we obtain a simple proportionality
relation
between the variations of the spinorial phase angle and of the zbw-plane
rotation angle:
$$
\Delta\varphi \ug {1\over 2}\Delta\vartheta \; .
\eqno{(20)}
$$
{\em In such a way, the so-called} U(1) {\em gauge transformations get a
straightforward and clear
geometrico--kinematical meaning}. \ They indeed may be regarded not just as
rotations in an
abstract space (namely, the Gauss plane of the complex
spinors), but actually
as spatial rotations in the {\em physical} spin plane.
Thus the {\em electromagnetic gauge invariance} ---global or local as it be---
owned by the currents,
the wave-equations, and the interaction lagrangians, {\em means}, as a matter
of fact, {\em that  currents, energies and forces are indepemdent of
the instantaneous angular position of the point-like charge}.

\

\h D) {\em The total angular momentum conservation:}

\

\cent{$(\Om\wedge S)\cos \bt = {\dot S}\cos \bt = p\wedge v$}

\

\noindent (the Lagrange equation obtained by the variation with respect
to $R_0$ has been
multiplied on the left by $R_0$, so that it has been singled out the
2-vectorial part).

\

\h In our Hamilton--Jacobi equation  (case A)) it {\em does not appear a
``quantum
potential''}, which is quite present, on the contrary, in the Dirac
fluid.$^{[10]}$   Moreover,
the total angular momentum {\em is locally conserved}, whilst this is not
the case for the
Dirac fluid. Thus, in conclusion, we can state that the present
hydrodynamics is that of a typical Weyssenhoff fluid.\\

\section{Operations on spinors and rotations in the spin plane}

In this section we are going to point out the interesting interrelation
occurring between some important transformations acting on spinors
and the orbital zbw motion of the electric pointlike charge $\cal Q$ in the
spin plane. \  In the previous section, we discussed
about the remarkable correlation between the gauge transformation U(1) and a
general rotation on the spin plan; \
in what follows we shall deal with the two-valuedness nature of the fermionic
wave-function, and with the parity and charge conjugation transformations.

\

As is well-known, in the standard framework of the quantum wave-mechanics the
sign of the fermion wave-function ---at variance with the scalar
particles case--- does change if we make a 360$^{\rm o}$-rotation
of the reference frame around an arbitrary axis. With regard to this,
we can really get a quite simple and natural classical interpretation
in the framework of the present BZ model. \
Without any recourse to fibre-bundles or to other topological tools,
we shall succeed in understanding why the phase of the quantum final state
varies even if the final particle position remains unchanged. \
As seen above, in the BZ model the phase of the wave-function is strictly connected
to a geometric-kinematical quantity: the phase angle of the position of
the internal ``constituent" {\cal Q} in the spin plane.
Now, a 360$^{\rm o}$ rotation around the $z$-axis of the coordinate frame 
(``passive
point of view'') is fully equivalent to a 360$^{\rm o}$ rotation, around the 
same axis,
of our microsystem, and therefore of the rotating charge (``active point of
view''). \
On the other hand, as a consequence of the last transformation, the zbw angle
$2m\tau$ in  $x(\tau)$, eq.(5c), suffers a variation of 360$^{\rm o}$ and the
proper time $\tau$ increases of a zbw period $T_{zbw}=\pi/m$, that is just what
happens when {\cal Q} performs a complete circular orbit around the $z$-axis.
But, because the period $T_{\psi}=2\pi/m$ of the spinor $\psi(\tau)$ in (5a)
is twice the period $T_{zbw}=\pi/m$ of the zbw motion, such a spinor {\em
results to suffer a phase increment of only {\rm 180}$^{\rm o}$}, and then changes
sign (e$^{i\pi} \equiv -1$) so as it does in standard quantum mechanics.

\

Analogous considerations can be made in connection with the {\em parity
transformation}. In this case the ``active'' operation consists in the specular
riflession of {\cal Q} around the origin of the cartesian axes.
In the CMF it is equivalent, once the chosen motion--plane is the $xy$-plane, to
to a 180$^{\rm o}$-rotation of $\cal Q$ around the $z$-axis. Once again, a
180$^{\rm o}$--rotation of $\cal Q$ implies a
spinor-phase variation of just a half of it 
(i.e., a 90$^{\rm o}$--variation), because, as seen, we have
$T_{\psi}=2T_{zbw}$. Now, if we take $m\tau=\pi/2$ in the spinor solution (5a),
we immediately get
$$
\psi(\pi/2) = -i\ga_0\psi(0) \; .
\eqno{(21)}
$$
Let us recall$^{[14]}$, at this point, that the {\em parity operator} in
Dirac wave-mechanics is nothing but $\Po = \pm i\ga_0$ (the choice of the
sign being actually arbitrary). In this way, once again, the formal features
of such a quantum operator get a straight meaning in the classical context
of the BZ model. The intrinsic non-intuitive property owned by the electron
state vector,
for which the double application of a parity operation is {\em not} an identity
---indeed we have $\Po^2\equiv-\1$---, is now related to the peculiar fact 
that parity 
really corresponds to a 90$^{\rm o}$ rotation of $\cal Q$.

\

Finally, let us consider the basic transformation of the relativistic quantum
mechanics: charge conjugation.
If we, as usual, associate the negative energies (and then assume in the 
present SCM frame, $m<o$) to antiparticles, we shall have, with regard to the particles
case, a simple {\em inversion of the rotation direction}, 
the other motion kinematical features remaining unchanged.  
All this comes out immediately from the
motion equation (5c), where only the sign of the odd function $\sin (2m\tau)$
changes when we make the $m\longra -m$ transformation, whilst the even function
$\cos (2m\tau)$ is left unchanged. Therefore for free polarized particles$^{[9]}$
condition $s_z = +\frac{1}{2}$
does imply anti-clockwise and clockwise circular uniform motions for electrons and
positrons, rispectively. \
This result seems to agree with the interpretation of the antiparticle states as
``time-inverted'' states (with reversed--sign energy too!$^{[15]}$), for 
which the ``time arrow points in the direction opposed
to the one of particles''. For such an interpretation within the 
classical contexts see refs.$^{[15]}$ and refs. therein.

\h Furthermore, analogously to what seen above, it is possible in the present
approach to forward a simple
classical deduction of {\em relative fermion-antifermion parity $P_r$}, which
is known to be equal to -1. In fact, the phase of the state vector for the
electron-positron system, which is a factorization of the electron and positron
wave-functions, suffers a total variation of 180$^{\rm o}$. This because, while the
particle state vector, under parity, results rotated of a +90$^{\rm o}$-angle, the antiparticle
vector is instead rotated of the same magnitude, but in the opposit direction,
that is of a -90$^{\rm o}$-angle.

\vs{2.0cm}

{\bf Acknowledgements}

This article is dedicated to the memory of Asim O. Barut. \ The authors wish
to acknowledge continuous, stimulating discussions with  H.E. Hern\'andez,
W.A. Rodrigues Jr., J. Vaz and D. Wisnivesky. \
Thanks for useful discussions and kind collaboration are also due
to G. Andronico, G.G.N. Angilella, M. Borrometi, L. D'Amico,
G. Dimartino, C. Dipietro, P. Falsaperla, A. Lamagna,
L. Lo Monaco, E. Majorana Jr., R.L. Monaco, E.C. de Oliveira, 
M. Pav\v{s}i\v{c}, R. Pucci,
M. Sambataro, S. Sambataro, R.Turrisi and M.T. Vasconselos.

\end{document}